\documentclass[graybox]{svmult}

\usepackage{mathptmx}       % selects Times Roman as basic font
\usepackage{helvet}         % selects Helvetica as sans-serif font
\usepackage{courier}        % selects Courier as typewriter font
\usepackage{makeidx}         % allows index generation
\usepackage{graphicx}        % standard LaTeX graphics tool
\usepackage{multicol}        % used for the two-column index
\usepackage[bottom]{footmisc}% places footnotes at page bottom

\makeindex             % used for the subject index
\newcommand{\lsi}{LS~I~+61$^\circ$303}
\begin{document}

\title*{Fermi results on $\gamma$-ray binaries}
\author{A.~B. Hill \and R. Dubois$^{\star}$ \and D. Torres \\ on behalf of the \textit{Fermi}-LAT collaboration}
\authorrunning{A.~B. Hill \and R. Dubois \and D. Torres}

\institute{A.~B. Hill \at Universit\'e Joseph Fourier, LAOG, UMR 5571, BP 53, 38041 Grenoble Cedex 09, France \\ \email{adam.hill@obs.ujf-grenoble.fr}
\and R. Dubois ($^{\star}$ speaker) \at W. W. Hansen Experimental Physics Laboratory, KIPAC, Department of Physics and SLAC National Accelerator Laboratory, Stanford University, Stanford, CA 94305, USA \\ \email{richard@slac.stanford.edu}
\and D. Torres\at Instituci\'o Catalana de Recerca i Estudis Avan\c{c}ats, Barcelona, Spain \\ \email{dtorres@ieec.uab.es}}
\maketitle

\abstract*{The past decade has witnessed a revolution in the field of observational high energy $\gamma$-ray astrophysics with the advent of a new generation in ground-based TeV telescopes and subsequent GeV space telescopes.  The \emph{Fermi} Large Area Telescope (LAT) was launched in August 2008 and has offered unprecedented sensitivity and survey capabilities in the 30 MeV - 300 GeV energy range.\newline\indent
Presented here are the results from the first two years of LAT observations of galactic binary systems including the definitive detections of \lsi, LS 5039 and Cyg X-3.  These sources and others are discussed in context with their known TeV and X-ray properties.  The LAT data provide new understandings and pose new questions about the nature of these objects.  The identification of an exponential cutoff in the spectra of both \lsi\ and LS 5039 was unexpected and poses challenges for explaining the emission mechanisms and processes which are in operation within these systems.}

\abstract{The past decade has presented a revolution in the field of observational high energy $\gamma$-ray astrophysics with the advent of a new generation in ground-based TeV telescopes and subsequent GeV space telescopes.  The \emph{Fermi} Large Area Telescope (LAT) was launched in August 2008 and has offered unprecedented sensitivity and survey capabilities in the 30 MeV - 300 GeV energy range.\newline\indent
Presented here are the results from the first two years of LAT observations of galactic binary systems including the definitive detections of \lsi, LS 5039 and Cyg X-3.  These sources and others are discussed in context with their known TeV and X-ray properties.  The LAT data provides new understandings and pose new questions about the nature of these objects.  The identification of an exponential cutoff in the spectra of both \lsi\ and LS 5039 was unexpected and poses challenges for explaining the emission mechanisms and processes which are in operation within these systems.}

\pagebreak
\section{Introduction}
\label{sec:1}
To date there are only five X-ray binaries that have reported detections of high energy (HE; 0.1-100 GeV) or very high energy (VHE; $>$100 GeV) $\gamma$-ray emission: \lsi~\cite{2006Sci...312.1771A,2009ApJ...701L.123A}; LS 5039~\cite{2005Sci...309..746A, 2009ApJ...706L..56A}; PSR B1259$-$63~\cite{2005A&A...442....1A}; Cyg X-3~\cite{1988PhR...170..325B,2009Sci...326.1512F}; Cyg X-1~\cite{2007ApJ...665L..51A, 2010ApJ...712L..10S}.  Whilst there were theoretical predictions that these systems may be capable of emitting such high energy radiation and some reports of high energy sources that may be associated with some of the above systems the first definitive detections and identifications have only occurred within the past five years due to the current generation of sensitive HE and VHE telescope facilities.

PSR B1259$-$63 was long known to be the only millisecond pulsar in orbit around a main-sequence star. It was a known hard X-ray source and was predicted to produce $\gamma$-rays through the interaction of the pulsar wind with the disk of the Be companion; VHE radiation was discovered by the H.E.S.S. experiment~\cite{2005A&A...442....1A} confirming this theoretical prediction. \lsi\ and LS 5039 are both persistent, well known X-ray binaries which were spatially consistent with HE sources detected by the Energetic Gamma-Ray Experiment (EGRET) in the 1990s~\cite{1999ApJS..123...79H}, however the lack of detection of periodic flux modulation and the size of the error box meant it was impossible to definitively claim a detection of $\gamma$-rays from these sources.  This all changed with first the detection of these sources at VHE
by the MAGIC, VERITAS and H.E.S.S. Cherenkov radiation telescopes and subsequently the detections at HE by the \emph{Fermi} Gamma-Ray Space Telescope.  Cyg X-1 and Cyg X-3 are examples of microquasars, accreting black holes or neutron stars in binary systems which have associated relativistic jets.  A key tool for identifying binary systems is their orbital modulation, due to the regular pattern of emission as the orbit is traversed.  Typically the zero phase of the orbit may be assigned from radio data, the time of periastron or some other marker, but is arbitrary.  Until 2009, there had been no unambiguous detection of $\gamma$-ray emission from a microquasar system.  The identification became conclusive with the confirmation of modulation at the orbital period.

We present here the latest results on the HE and VHE $\gamma$-ray emission from this population of sources, with an emphasis on the results and discoveries from the \emph{Fermi}-LAT.

\section{LS I +61 303}

\subsection{The original discovery, and further TeV observations:
flux, spectrum, periodicity}

The very-high energy (VHE) spectrum derived from MAGIC data between
~200 GeV and ~4 TeV at orbital phases between 0.4 and 0.7 was fitted by a power law
function: $ F_\gamma = (2.7 \pm 0.4 \pm 0.8) \times 10^{-12}
(E/{\rm TeV})^{-2.6 \pm 0.2 \pm 0.2} \; {\rm cm}^{-2} {\rm s}^{-1}
{\rm  TeV}^{-1}, $ with the errors quoted being statistical and
systematic, respectively~\cite{2006Sci...312.1771A}. All further reports of
\lsi\ at very-high energies would produce compatible spectra with this
measurement, within errors. These first MAGIC measurements
showed that the very-high energy $\gamma$-ray emission from LS I
+61 303 was variable. The maximum flux corresponded to about 16\%
of that of the Crab Nebula, and was detected around phase 0.6 with
a significance of 8.7$\sigma$. The source is no longer detected near periastron
and only upper limits were therein imposed.
The VERITAS array carried out independent observations and soon confirmed
these results~\cite{2008ApJ...679.1427A}. These early MAGIC/VERITAS observations occured at a number of different orbital phases but the source was only ever detected at similar phases; this hinted at a periodic nature of the emission.

Further MAGIC observations showed that this emission is indeed
periodic~\cite{2009ApJ...693..303A}, with the system showing regular
outbursts at TeV energies in phases around 0.65 and no significant
signal elsewhere.
The periodicity was tested using the Lomb-Scargle perodogram both on a set of OFF
data and on the source of choice. Whereas the former produced no
periodicity, the periodogram analysis of the latter did.
The maximum peak in the periodogram is seen at frequency
$\nu=0.0373$d$^{-1}$ which had a post-trial chance probability of
$2\times10^{-7} $. Other peaks (although less significant)
were seen in the signal analysis and not seen in the background sample.
When folding the data with the frequency obtained, the data did not seem
at first sight to be described by a simple sinusoidal shape (in part,
perhaps due to spacing and quality). In fact, fitting a sine function
and subtracting it from the power analysis removes the peak related to
the orbital period but not the others, something that is improved when
a sine plus a gaussian function is fitted to the data, and subsequently
subtracted from the analysis.
Thus, whereas further MAGIC observations succeeded in bracketing the
periodicity of \lsi\ in TeV  $\gamma$-rays to be very close to the
orbital period (26.8$\pm$0.2)~days, as well as to other wavelength
estimates (e.g., at 1.5$\sigma$ from the best estimate, coming from
radio of 26.4960$\pm$0.0028 days~\cite{2002ApJ...575..427G}), with the error determined by Monte-Carlo
simulations, they also reveal a new mystery for the flux evolution of
\lsi\ at these energies.
Is there an average (sinusoidal) emission along the orbit,
plus an extra component associated with the turn on of a different
process (the Gaussian) that only appears at the TeV maximum phase
range?

\subsection{TeV and X-ray simultaneous observations}

The first {\it strictly  simultaneous} observations at TeV and X-rays
of \lsi\ have been recently presented by the MAGIC collaboration~\cite{2009arXiv0907.0992J}. Using observations by  {\it
XMM-Newton}, \textit{Swift}, and MAGIC during $60\%$ of an orbit, it was found
that there is a correlation between the X-ray and TeV emission at the time of the
TeV maximum.
A linear fit to the six MAGIC/{\it XMM-Newton} pairs of observations that
trace the outburst yielded a correlation coefficient of $r=0.97$. A
linear fit to all ten simultaneous pairs (including \textit{Swift} data) provides a
high correlation coefficient of $r=0.81_{-0.21}^{+0.06}$ (which has a
probability of about $5\times10^{-3}$ to be produced from
independent X-ray and TeV fluxes). Due to strong variability of the
X-ray emission of \lsi, of about 25\% in hour scales, other previous
campaigns with contemporaneous, but not strictly simultaneous data, as
for instance the previous MAGIC campaign~\cite{2008ApJ...684.1351A}.
2008, or the VERITAS campaign~\cite{2009ApJ...700.1034A} come
short in reaching such a result, which, nevertheless, would not suffer
from adding further statistics and confirmation. Therefore, the
X-ray/TeV correlation found for \lsi\ points to emitting processes happening at the same time in both
wavelengths, and being the result of the same physical population of
particles; especially, since at the TeV
maximum one would not expect the TeV photons to be subject to
significant absorption due to gamma-gamma processes.

\subsection{The {\it Fermi} results on \lsi}

With the launch of the {\it Fermi} Gamma-ray Space Telescope on
11 June 2008 and the beginning of operations in its survey mode~\cite{2009ApJ...697.1071A}, coverage of the full sky above 100 MeV is available through its main instrument, the Large Area Telescope (LAT).
There are two main results to extract out of the {\it Fermi}
observations of \lsi\ using the data collected in the first 9 months
after launch~\cite{2009ApJ...701L.123A}:
\begin{enumerate}
	\item The unambiguous detection of
a source co-located with the position of \lsi\, which is seen to be
periodic (too) at GeV energies (the first such periodicity ever
detected) and anti-correlated with the TeV emission.
	\item The discovery of a cutoff in the spectrum of the {\it Fermi} (GeV) source.
\end{enumerate}

\begin{figure}[t]
%  \begin{flushright}
   \includegraphics[width=.4\linewidth,angle=-90.]{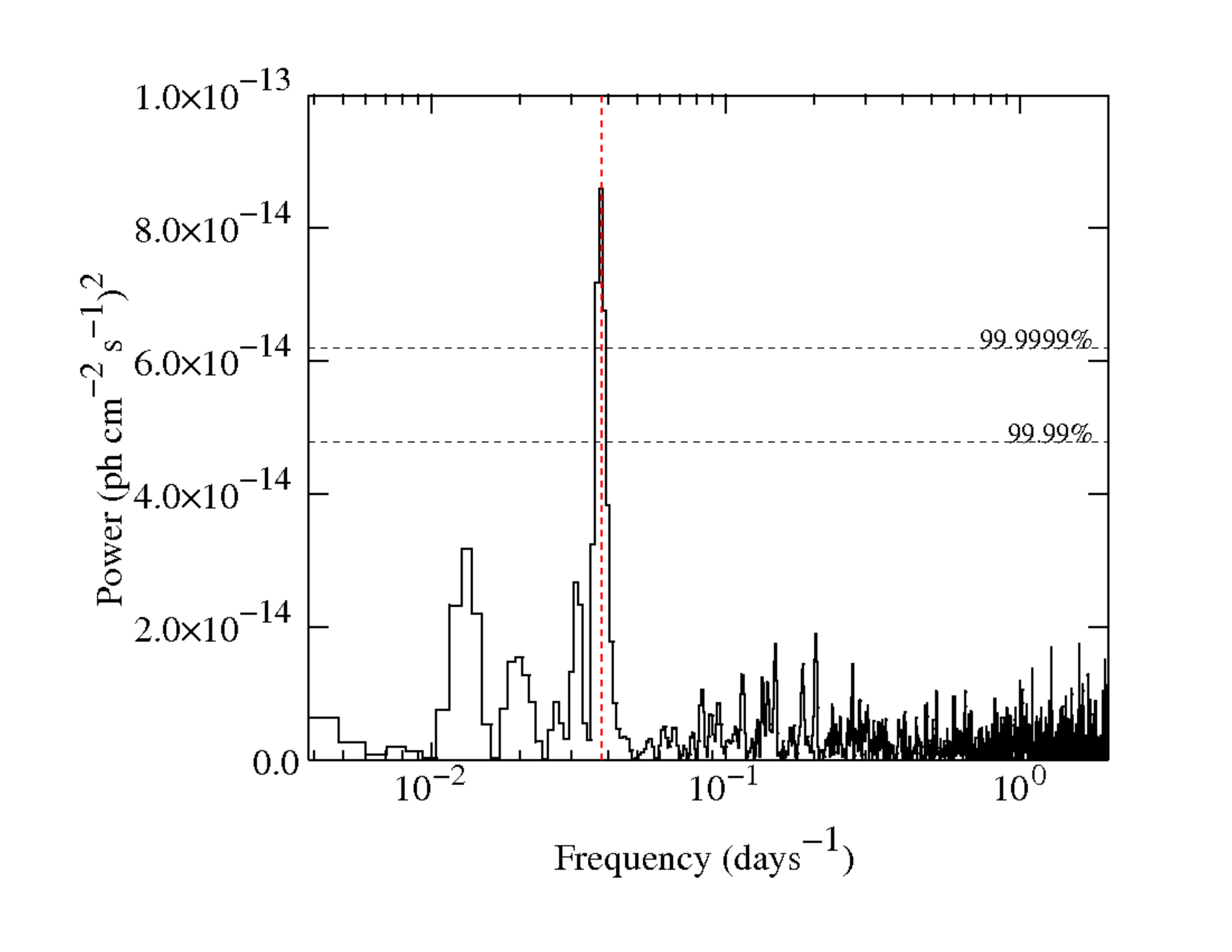}
   \hspace{-0.8cm}
     \includegraphics[width=.4\linewidth,angle=-90.]{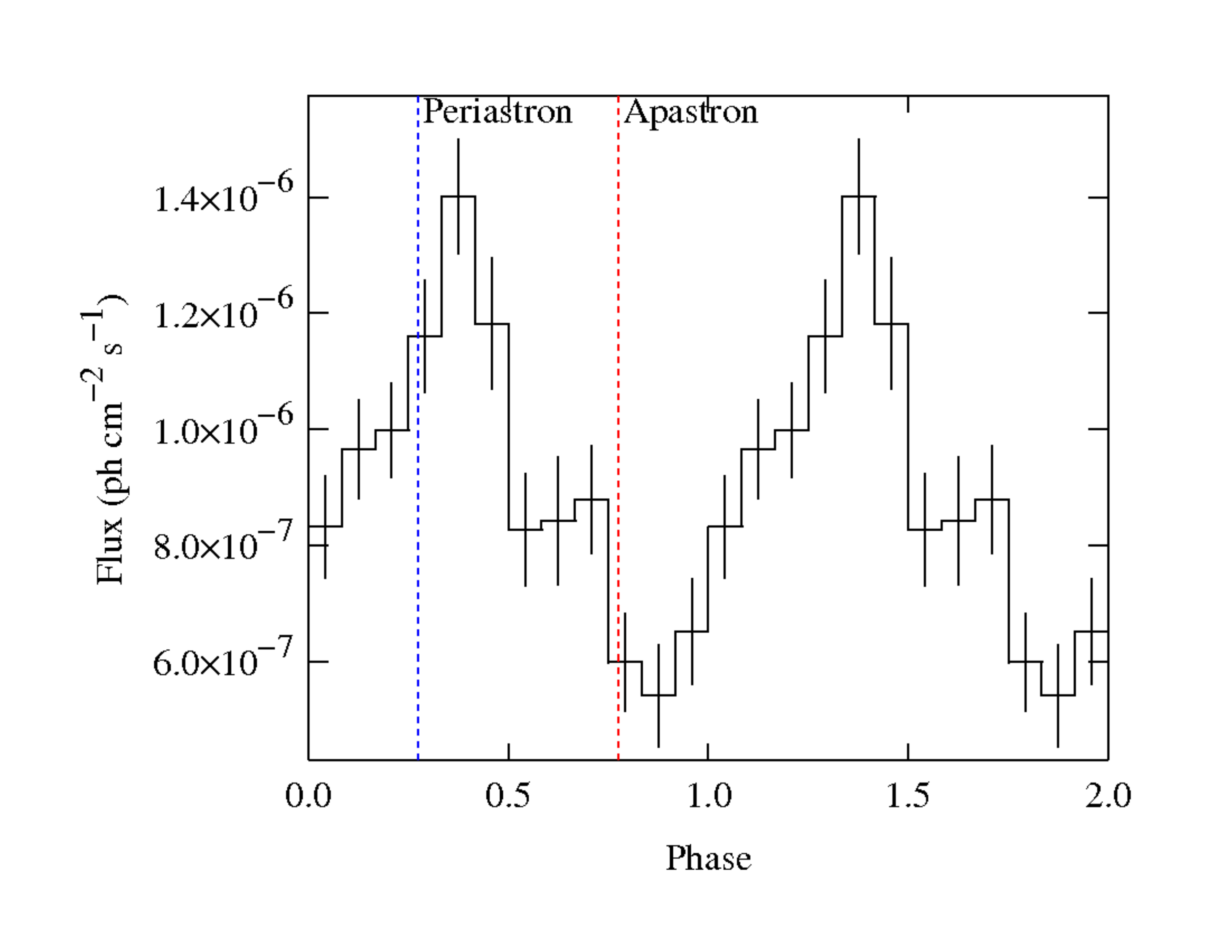}
%    \end{flushright}
  \caption{ The power spectrum (left) and phase folded light curve (right) of \lsi\
at 100 MeV -- 20 GeV. Left: The weighted Lomb-Scargle periodogram of the LAT light curve. The vertical dashed line indicates the known orbital period~\cite{2002ApJ...575..427G}; the horizontal dashed lines indicate the marked significance levels.
Right: The phase folded 100 MeV -- 20 GeV light curve of \lsi. The dashed lines indicate periastron and apastron of the system.}
  \label{fermiLSIpow}
\end{figure}

\lsi\ is detected at significance of $\sim$70$\sigma$ at a position consistent with the optical location of the source. The LAT light curve was constructed using aperture photometry of 100 MeV -- 20 GeV events.  From a 3.2 hour binned light curve the weighted Lomb-Scargle periodogram~\cite{2010arXiv1001.4718C} was calculated.  The periodogram exhibits a clear peak at 26.6 $\pm$ 0.5 days (1$\sigma$ errors obtained though Monte-Carlo) consistent with the known orbital period. 
The binned LAT light curve folded on the nominal orbital
period~\cite{2002ApJ...575..427G} with the usual zero phase at ${\rm MJD}~43,366.2749$~\cite{1979AJ.....84.1030G} shows a large modulation amplitude with maximum
flux occurring slightly after periastron passage (where there is no
TeV emission detected, neither by MAGIC nor by VERITAS).  The folded light curve and the Lomb-Scargle periodogram are shown in Figure~\ref{fermiLSIpow}.
The overall light
curve can be reasonably well fit by a simple sine wave. Fitting a sine
wave to each of the individual 9 orbits observed finds that the best fit
amplitude varies between 6.8$\pm$0.9 and 2.2$\pm$0.9 $\times$10$^{-7}$ photons
cm$^{-2}$ s$^{-1}$~\cite{2009ApJ...701L.123A}, which might be indicative of some orbit-to-orbit
variability.

The LAT data was well fit by a power-law plus exponential cutoff returns; the chance probability that a simple power-law was incorrectly rejected is 1.1$\times10^{-9}$.  The photon index was found to be $\Gamma=$2.21 $\pm$
0.04 (stat) $\pm$ 0.06 (syst); the flux above 100 MeV is (0.82 $\pm$
0.03 (stat) $\pm$ 0.07 (syst)) $\times 10^{-6} $ photons cm$^{-2}$ s$^{-1}$
and the cutoff energy is 6.3 $\pm$ 1.1 (stat) $\pm$ 0.4 (syst) GeV.
Reduced statistics in a phase-bin of 0.1 of the orbit precluded us from
finding a statistically significant cutoff in each of the bins, and so a
simple power law was fitted. No significant variation of this photon
index with phase was found.

During part of the {\it Fermi} measurements, there was a VERITAS
campaign~\cite{2009arXiv0907.3921H} observing at higher energies (100 GeV -- 30 TeV), and no
detection was found even when part of the TeV maximum phase region was
covered. The lack of strong emission from \lsi\ during
these observations is at face value somewhat surprising, especially if
one looks at the light curve in the orbit where all are upper limits,
but does not
contradict (yet) previous measurements. The TeV maximum has
happened around phase 0.6 with excursions down and up in phases of about 0.2
of the orbit. The apastron coverage during this campaign was limited,
with large data gaps, e.g., the coverage between phases 0.72 and 0.82 consists of only 40 minutes.
Consequently the low statistics of the VERITAS measurements during this campaign do not allow firm conclusions to be drawn regarding the disappearance of the signal or on its deviation from the usual periodic behavior.
It is clear, however, that any future ground-based \lsi\ campaign will
benefit from automatic {\it Fermi} coverage. At the same time, a
strictly periodic behavior would be easy to rule out by subsequent
MAGIC/VERITAS observations of the TeV maximum, and the problem of
understanding the origin of the  $\gamma$-ray  emission would be even
more complex if this were so.

\begin{figure}[t]
%  \begin{flushright}
   \includegraphics[width=.5\linewidth]{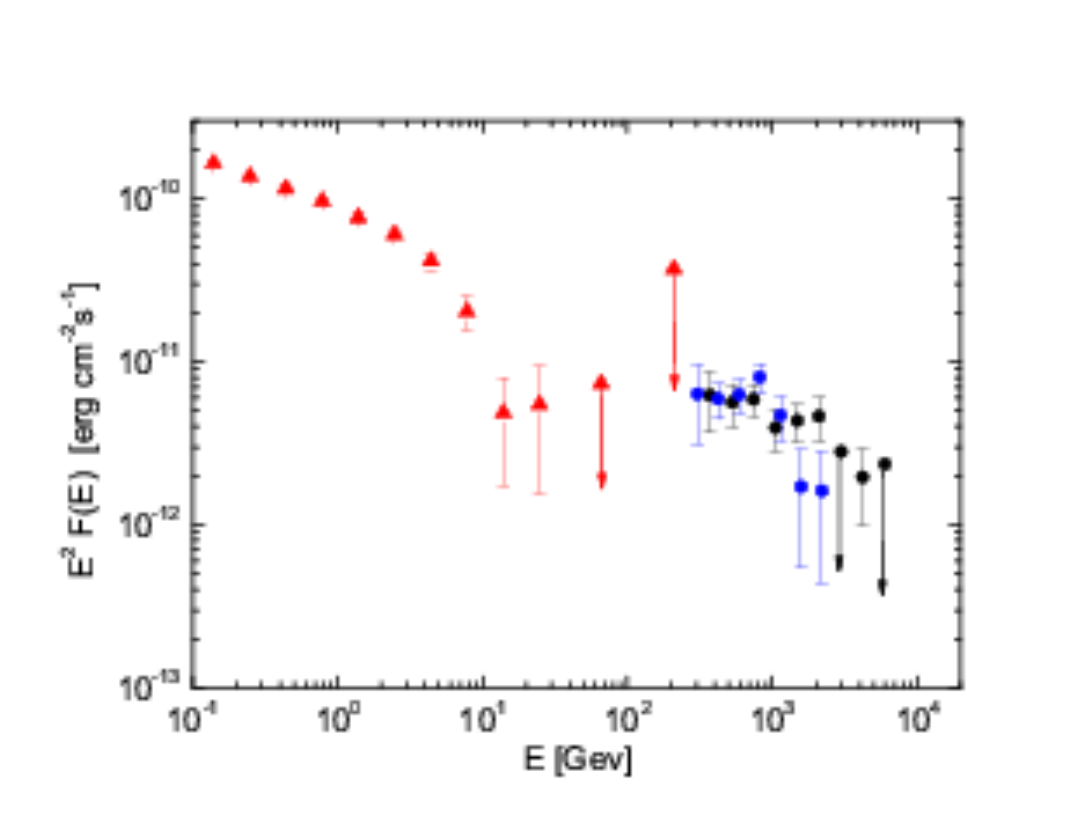}
   \hspace{-0.8cm}
     \includegraphics[width=.5\linewidth]{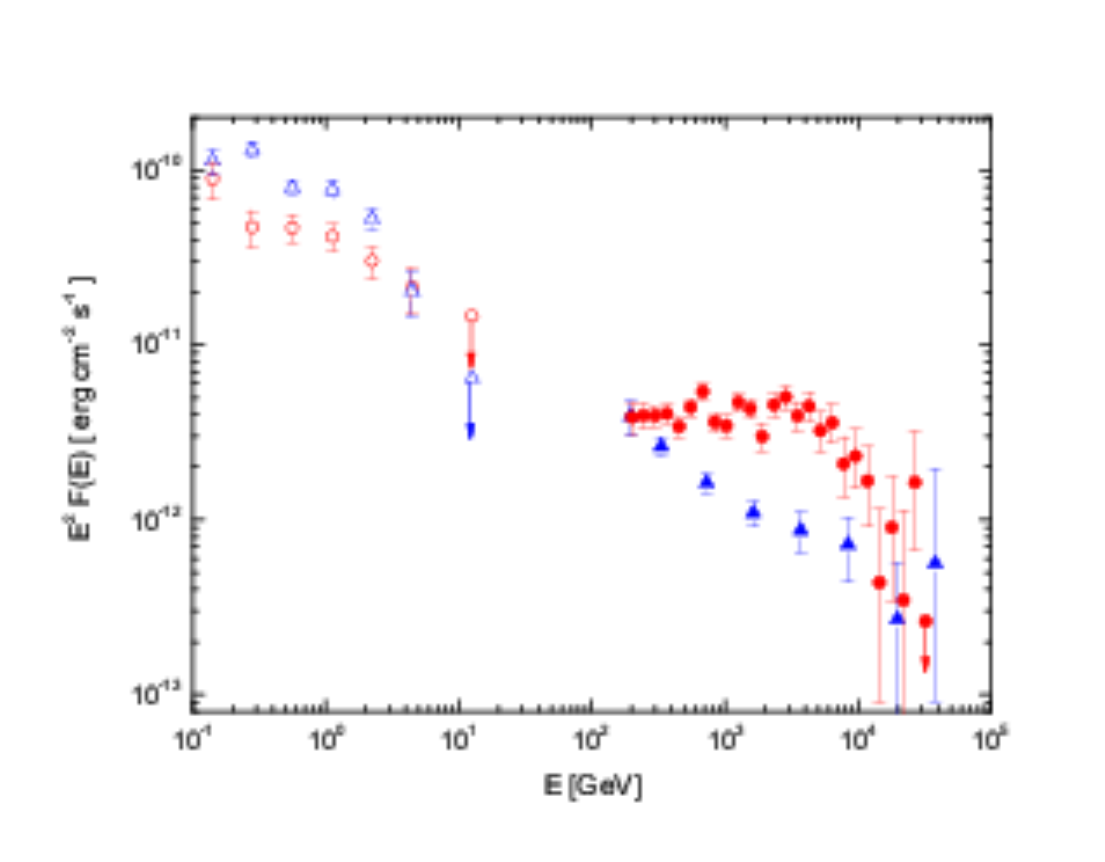}
%    \end{flushright}
  \caption{ High and very-high energy spectral data points of \lsi\
(left) and LS 5039 (right). Left: {\it Fermi} data points are red;
MAGIC data are blue  (high state phases 0.5-0.7); VERITAS data are
black (0.5-0.8).
Right: The red data points show the spectrum at
inferior conjunction (TeV maximum, GeV minimum, phases 0.45--0.9; the blue data points
show the spectrum at superior conjunction
(TeV minimum, GeV maximum, phases $<0.45$ and $>0.9$). Higher energy
data come from H.E.S.S.
Note that the data from the different telescopes are not
contemporaneous, though they do cover multiple orbital periods.}
  \label{fermiLSspectrum}
\end{figure}

%%%%%%%%%%%%%%%%%
\section{LS 5039}

\subsection{The original discovery, and further TeV observations:
flux, spectrum, periodicity}

LS 5039 was discovered by the H.E.S.S. experiment to be a
$\gamma$-ray  source\cite{2005Sci...309..746A} and soon afterwards,
periodicity in the TeV $\gamma$-ray flux, consistent with the known orbital
period~\cite{2005MNRAS.364..899C}, was also reported~\cite{2006A&A...460..743A}.
Unlike \lsi\, H.E.S.S. observations of LS 5039 have detected this source at all orbital phases observed.
The authors of \cite{2006A&A...460..743A} define two broad phase intervals, the inferior
conjunction ($0.45<\varphi<0.9$) and superior conjunction ( $\varphi>0.9$ and
$\varphi<0.45$), for which different (averaged) H.E.S.S. spectra were produced.
The differential photon energy spectrum ($0.2$ to $10.0$ TeV) for inferior conjunction
is consistent with a hard power-law with an exponential cutoff, where
the slope is $\Gamma = 1.85 \pm 0.06$ (stat) $\pm$ 0.1 (syst) and the
exponential cutoff energy is at $E_0 = 8.7 \pm 2.0$  TeV (for the
fitted function $dN/dE \sim E^\Gamma  \exp(-E/E_0))$. In contrast, the
spectrum for superior conjunction is consistent with a relatively steeper ($\Gamma=
2.53 \pm 0.07$ (stat) $\pm$ 0.1 (syst)) pure power-law (0.2 to 10
TeV).
The statistics collected along the orbit actually allowed to find that
the parameters of power-law fits to the $\gamma$-ray data obtained in 0.1 phase binning already
displayed significant variability. Because of low statistics, and presumably too, of different
statistics at the higher end of the spectrum in each of these bins,
more complicated functions such as a power-law with
exponential cutoff provided no better fit than that of a pure power-law. These results, together with the
H.E.S.S. light curve and averaged spectra are shown below, where we make a 
comparison with various models.

\subsection{The {\it Fermi} results on LS 5039}

\begin{figure}[t]
   \includegraphics[width=.9\linewidth]{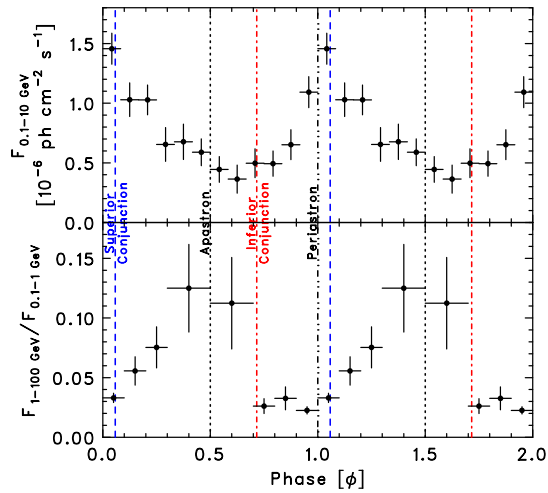}
  \caption{ \textit{Fermi} average phase folded light curves for LS 5039. Top: 0.1--10 GeV flux variations with orbital phase. Bottom: Changes in the hardness ratio, flux(1--100 GeV)/flux(0.1--1 GeV), across the orbit. }
  \label{fermiLS5039-LC}
\end{figure}

The dataset for the analysis reported by {\it Fermi}
spanned the period from 4 August 2008, through 29 April 2009, and thus
covered multiple orbits of the system~\cite{2009ApJ...706L..56A}.
Like \lsi\ this system is also found to be periodic in GeV $\gamma$-rays (3.903$\pm$0.005
days) and anti-correlated with the TeV emission. The periodicity is
thus also consistent with the orbital period. The SED also presents a
cutoff at a few GeV and is well fitted by a similar function:
$E^{-\Gamma} \exp(-E/E_{\rm cutoff})$, with
the photon index being $\Gamma=1.9 \pm$ 0.1(stat) $\pm$ 0.3 (syst); a
100 MeV--300 GeV flux of (4.9 $\pm$ 0.5 (stat) $\pm$ 1.8 (syst)) $\times
10^{-7}$ photons cm$^{-2}$ s$^{-1}$, and a cutoff energy at 2.1 $\pm$
0.3 (stat) $\pm$ 1.1 (syst) GeV.  The energy spectra of both LS 5039 and \lsi\ are shown in Figure~\ref{fermiLSspectrum}.
%Figure \ref{fermiLSspectrum} shows these results, and
{\it Fermi} observations were able to further separate into superior
and inferior parts of the orbit, in phase ranges as defined by
H.E.S.S.: GeV observations also pinpoint this (GeV / TeV --
anti-correlated) flux variability.
%Color coding in this figure is compatible with that used for higher energies.
The fact that H.E.S.S. did not find
long-term variability makes it reasonable to directly compare
non-simultaneous H.E.S.S. and {\it Fermi} observations.
Using the same phase intervals, the inferior conjunction slope was found to be $\Gamma=
2.25\pm0.11$ with no exponential cutoff energy required to provide a
better fit; whereas at superior conjunction, a $\Gamma=1.91\pm 0.16$ with a cutoff
energy of $1.9\pm0.5$ GeV was found. The spectral shape in {\it Fermi}
is softer around periastron (near superior conjunction) and is harder around apastron as shown in Figure~\ref{fermiLS5039-LC}.

\section{Cygnus X-3}
\subsection{Historical observations at high energies}
Cyg X-3 is a well known, powerful high-mass X-ray binary with a short orbital period of 4.8 hours~\cite{1988PhR...170..325B}.  The system comprises of a compact object, the nature of which is still subject to debate, accreting matter from a Wolf-Rayet companion star~\cite{1992Natur.355..703V}. The source was discovered in 1966 and has been widely observed across the electromagnetic spectrum. It regularly becomes the brightest radio source amongst the known binary systems with large flares attributed to its relativistic jets. Emission had been detected up to $\sim$300 keV whilst the system shows a complex X-ray spectrum which transitions between two main states; `soft' and `hard', the source is known to flare in radio when entering the `soft' state with associated relativistic plasma ejection events~\cite{2006MNRAS.369..603F}.

Historically there were reported detections at MeV--PeV energies in the 1970s and 1980s. These observations came from 30 MeV -- 5 GeV sensitive balloon and satellite borne experiments such as SAS-2 and COS-B, from atmospheric Cherenkov experiments at 0.1--500 TeV and extensive air shower experiments at 0.1--10 PeV~\cite{1988PhR...170..325B}.  However, the detections were typically of low significance and although some experiments claimed detections, others could not confirm them. These detections remained doubtful when the next generation of more sensitive ground based telescopes failed to confirm the TeV and PeV detection of Cyg X-3~\cite{1992ApJ...396..674O, 1997PhRvD..55.1714B} and the CGRO-EGRET failed to detect any GeV emission~\cite{1997ApJ...476..842M}. Hence, claims of HE and VHE emission from this microquasar were controversial and highly contested.

\begin{figure}
%  \begin{flushright}
   \includegraphics[width=.9\linewidth]{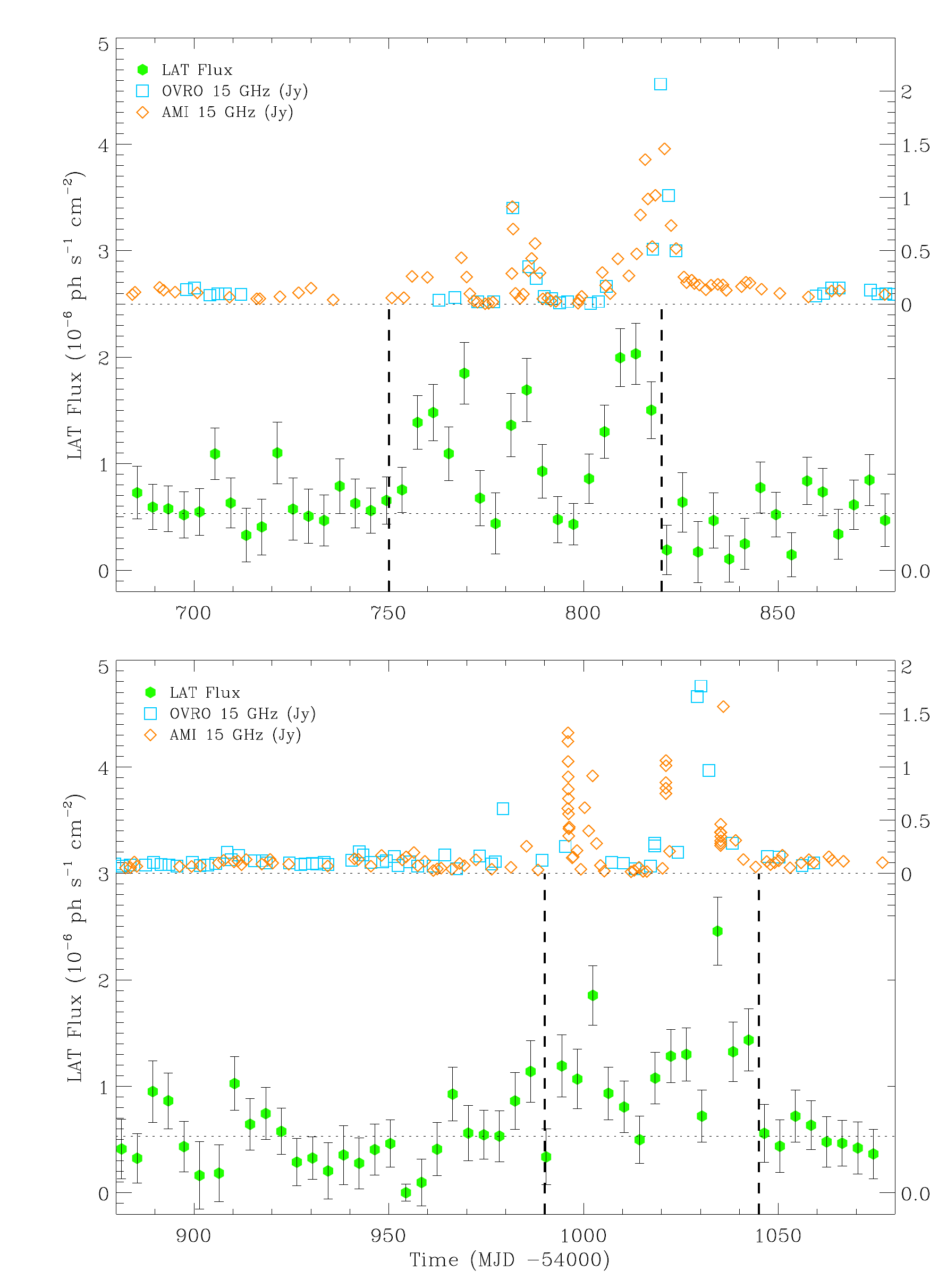}
%    \end{flushright}
  \caption{ Gamma-ray light curves of two of the flaring events observed by \emph{Fermi}.  The LAT fluxes $>$ 100 MeV (filled circles) are calculated from a likelihood analysis of 4-day long segments of data; a power-law with index of 2.7 is assumed in the analysis.  The vertical dashed lines indicate the periods the LAT detects flaring activity.  In the top of each panel the 15 GHz radio flux as measured by the AMI and OVRO 40-m radio telescopes is presented; an offset of 0.124 Jy has been removed from the OVRO data to compensate for the effect of extended nearby sources that are resolved by the AMI interferometer.  It is clear that a number of radio flares are detected coincident with the $\gamma$-ray activity detected by the LAT.}
  \label{CygX-3LC}
\end{figure}

\subsection{The Fermi results on Cygnus X-3}
The analysis of \emph{Fermi}-LAT observations of Cyg X-3~\cite{2009Sci...326.1512F} spans data taken from 4 August 2008 to 2 September 2009.  The Cygnus region is challenging to analyze due to the high levels of diffuse emission combined with the presence of 3 bright $\gamma$-ray pulsars~\cite{2010ApJS..187..460A}: PSR J2021+4026; PSR J2021+3651; PSR J2032+4127.  In fact, PSR J2032+4127, lies very close to the location of Cyg X-3 ($\sim$30$^\prime$) and contributes significantly to the large number of photons detected by the LAT at the location of Cyg X-3.  The impact of this pulsar was minimized by only using photons which arrive during the off-pulse phase. This preserves approximately 80\% of the live time and produces a significantly cleaner data set.

The LAT data yield a detection of a point source at the level of 29$\sigma$ at a location consistent with that of Cyg X-3.  The 4-day binned $\gamma$-ray light curve of the source flux $>$100 MeV (sections of which are shown in Figure~\ref{CygX-3LC}) clearly indicates that the source is highly variable.  The LAT detects Cyg X-3 during two specific active periods: 11 October to 20 December 2009 (MJD 54750--54820); 8 June to 2 August 2009 (MJD 54990--55045).  These active phases comprises one or more flares with a peak flux of up to $\sim$2 $\times$ 10$^{-6}$ photons cm$^{-2}$s$^{-1}$ above 100 MeV.  The average spectrum over the two active periods is represented by a single power-law model with a spectral index $\Gamma$ = 2.70 $\pm$ 0.05 (stat) $\pm$ 0.20 (syst) and an average flux above 100 MeV of [4.0 $\pm$ 0.3 (stat) $\pm$ 1.3 (syst)] $\times$ 10$^{-10}$ erg cm$^{-2}$s$^{-1}$.  Assuming a distance of 7 kpc this gives a luminosity, L$_{\gamma} \sim$ 3 $\times$ 10$^{36}$ erg s$^{-1}$, comparable to what was reported by SAS-2~\cite{1977ApJ...212L..63L}.  The \emph{AGILE} mission also reported detecting $\gamma$-ray activity coincident with the location of Cyg X-3 during these epochs as well as activity from 16-17 April 2008 (MJD 54572--54573), prior to the launch of \emph{Fermi}~\cite{2009Natur.462..620T}.  The \emph{AGILE} team report a 5.5$\sigma$ detection with a comparable average flaring flux to that reported by the \emph{Fermi}-LAT.

The association of the $\gamma$-ray flaring source was definitively confirmed though the detection of the 4.8 hour orbital period in a finely binned (1000s) aperture photometry light curve.  The periodicity was found using the Lomb-Scargle periodogram and weighting the input flux data points by their relative exposure.  No orbital periodicity is evident when using the entire data set, however restricting the data set to the periods of LAT detected enhanced emission discussed above yields a strong detection with a nominal false detection probability of 3.6$\times$10$^{-5}$.  Comparing the folded 0.1-100 GeV light curve with the 1.5--12 keV X-ray light curve taken from RXTE/ASM data shows that both have the same asymmetric shape with a slow rise and faster decay.  However, the LAT minimum trails the X-ray minimum by 0.3--0.4 in phase.

The relationship between the $\gamma$-ray and radio emission was confirmed through a discrete cross-correlation analysis which indicated a positive correlation between the two wavebands with a significance of $>3\sigma$.  The lag of the radio light curve to the $\gamma$-rays is not well constrained and is estimated to be 5 $\pm$ 7 days.  This specifically links the $\gamma$-ray activity to periods of relativistic ejection events.  The LAT flux is compatible with extrapolations of the hard X-ray tail observed above 30 keV, reportedly up to several 100 keV~\cite{2009MNRAS.392..251H}, in instances of the `ultrasoft' state.  In this case the \emph{Fermi} emission could be explained by inverse Compton scattering of UV photons from the WR star off of high energy electrons.  However this scenario requires that the emission region is not too close to the accretion disk otherwise the $\gamma$-ray emission will be absorbed through pair production on soft X-ray photons coming from the disk.  
	
If inverse Compton scattering is the dominant $\gamma$-ray production method then it is expected that the peak in the $\gamma$-rays would correspond to the time of superior conjunction when the electrons are seen behind the WR star and hence the energetic electrons directed towards the Earth undergo head-on collisions with the stellar UV photons.  This would correspond to the X-ray minimum assuming the X-ray modulation is produced through Compton scattering in the WR wind which is approximately what is observed in the average folded light curves.  A proposed model to explain the observed behavior of Cyg X-3 at high energies is that of a jet launched around a black hole oriented not too far from the line of sight, which interacts with the WR stellar wind to produce a shock at a distance of 1–-10 times the orbital separation from the system~\cite{2010MNRAS.404L..55D}.  This shock is the location where electrons are accelerated to GeV energies and upscatter star photons.

\begin{figure}[t]
   \includegraphics[width=.9\linewidth]{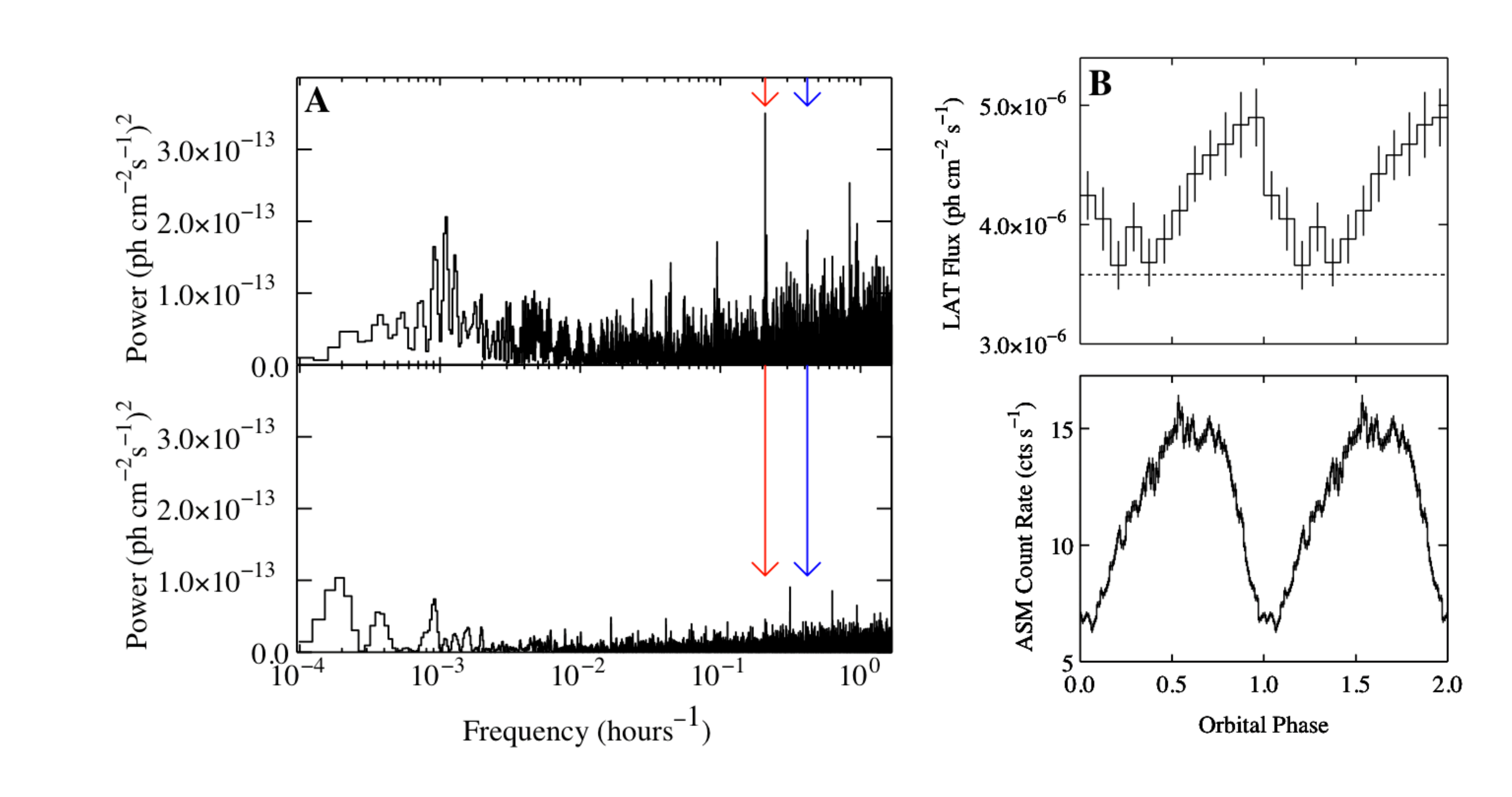}
  \caption{Left: The \textit{Fermi} power spectrum using only the data during the periods of enhanced emission (top) and for the entire LAT data set (bottom).  The vertical arrows indicate the frequencies of the orbital period and the second harmonic. \newline Right: The average orbital phase folded light curve of the LAT data during the periods of enhanced emission (top) and of the RXTE ASM X-ray data taking during the RXTE mission lifetime. Phase zero is set to be at the point of superior conjunction.}
  \label{fermiCygX3-LC}
\end{figure}

\section{Cygnus X-1}
Cyg X-1 is an exceptionally bright, well studied high-mass X-ray binary and is believed to host a stellar mass black hole.  It is located at a distance of $\sim$2.2 kpc and comprises of a O9.7 Iab donor star in a 5.6 day orbit with the compact object~\cite{2005MNRAS.358..851Z}. Cyg X-1 is another example of a microquasar, it is believed to be principally powered by accretion and is seen to transition between `soft' and `hard' X-ray spectral states.  The radio emission is observed to be relatively stable during the `hard' state except for rarely observed radio flares~\cite{2006MNRAS.369..603F} and is undetected in the `soft' state.

\subsection{Reports of high energy and very high energy emission}
Between June and November 2006 the MAGIC telescope observed Cyg X-1 for a total of 46.2 hours.  The MAGIC team searched the entire data set and found no significant persistent VHE $\gamma$-ray source at the location of Cyg X-1 and reported a 95\% flux upper limit in the 150 GeV -- 3 TeV band of the order of 1--5\% of the Crab~\cite{2007ApJ...665L..51A}.  Searching for faster time-varying signals within any given night resulted in the detection an excess of 4.9$\sigma$ (4.1$\sigma$ after correcting for trials) resulting from 79 minutes of effective on-time on September 24 2006 UTC 22:17 -- 23:41.  The observed excess is reported to be consistent with a point source at the location of Cyg X-1 and excludes a nearby radio nebula.  The spectrum was fitted by a power law
function: $ F_\gamma = (2.3 \pm 0.6) \times 10^{-12}
(E/{\rm TeV})^{-3.2 \pm 0.6} {\rm cm}^{-2} {\rm s}^{-1}
{\rm  TeV}^{-1}, $ with the errors quoted being statistical only; the systematic uncertainty is estimated to be 35\% on the overall flux normalization and 0.2 in the determination of the spectral index~\cite{2007ApJ...665L..51A}.

The TeV flare was coincident with a hard X-ray flare detected by \emph{INTEGRAL}, \emph{Swift}-BAT and \emph{RXTE}-ASM.  The TeV detection was observed at the rising edge of the first hard X-ray peak and the MAGIC non-detection the following night coincided with the decay of the second hard X-ray peak and hypothesize that the hard X-ray and $\gamma$-rays may be produced in regions linked by the collimated jet~\cite{2007ApJ...665L..51A}; these processes would have different physical timescales explaining the shift in time between the TeV and X-ray peaks.

Recently the \emph{AGILE} satellite has reported the detection of transient $\gamma$-ray activity from Cyg X-1 above 100 MeV~\cite{2010ApJ...712L..10S}.  They used $\sim$315 days of data from July 2007 -- mid October 2009 giving a net exposure of $\sim$13 Ms.  Performing a likelihood analysis over the entire integrated dataset does not yield a significant detection of Cyg X-1 and places a 2$\sigma$ upper limit on the $\gamma$-ray flux (0.1--3 GeV) of 3$\times$ 10$^{-8}$ photons cm$^{-2}$s$^{-1}$.  The \emph{AGILE} team searched for variability on day timescales and reported a single flaring episode corresponding to 15 October 2009 UTC 23:13:36 -- 16 October 2009 UTC 23:02:24, with a 5.3$\sigma$ pre-trial (4$\sigma$ post-trial) significance level.  The location of the flare is consistent with the location of Cyg X-1 and corresponds to a 0.1--3 GeV $\gamma$-ray flux of (2.32 $\pm$ 0.66) $\times$ 10$^{-6}$ photons cm$^{-2}$s$^{-1}$.  At the time of the flare Cyg X-1 was in the `hard' X-ray spectral state, however, unlike the instance of the MAGIC flaring event there was no simultaneous flare seen in the X-ray (or any other) waveband.  In March 2010 the \emph{AGILE} team reported new transient activity in the Cygnus region which was compatible with the Cyg X-1 location~\cite{ATel2512}.  They reported a 5$\sigma$ detection integrating from 24 March 2010 UTC 02:24 to 25 March 2010 UTC 01:01 with a corresponding flux above 100 MeV of (2.0 $\pm$ 0.9) $\times$ 10$^{-6}$ photons cm$^{-2}$s$^{-1}$.

A search of the \emph{Fermi}-LAT data for signs of the October 2009 flare yielded no significant detection.  Extracting data from the exact period reported by \emph{AGILE} the LAT places a 95\% upper limit on the $\gamma$-ray flux (0.1--3 GeV) of $\sim$4$\times$ 10$^{-7}$ photons cm$^{-2}$s$^{-1}$ (assuming a power-law spectral shape with photon index, $\Gamma$=2.2)~\cite{blog-CygX-1}.  Similarly the LAT did not detect any flaring emission during March 2010.

\section{Conclusion}

The first couple of years of \textit{Fermi}-LAT operations have been highly fruitful in expanding our understanding of the high energy $\gamma$-ray emission properties of Galactic $\gamma$-ray binaries.  In fact the availability of sensitive telescopes spanning the MeV--TeV energy range in the past decade have opened up this entire field of study and the LAT is the latest instrument to add to and expand this field of study.  

Gamma-ray binaries have now been definitively detected in the GeV energy range; both \lsi\ and LS 5039 are strong detections which have been identified through their spatial location and the clear signs of orbital modulation in the LAT observed flux.  The orbital modulation of the GeV flux was found to be anti-correlated with the modulation observed at TeV energies with the GeV flux peaking around periastron in both cases.  The identification of an exponential cut-off power-law shape to the spectrum of both sources was not expected.  It has been noted that the spectral shape observed in both $\gamma$-ray binaries is reminiscent of that seen in the large family of $\gamma$-ray pulsars now detected by the LAT~\cite{2009ApJ...706L..56A} (i.e. a hard power-law spectrum with an exponential cutoff at $\sim$2.5 GeV). It has been proposed that this may indicate that the emission in the \textit{Fermi} energy range from both \lsi\ and LS 5039 may therefore be magnetospheric in origin, however, this raises a new problem as it is not clear how such emission would be modulated by the orbital period of the binary. A clear indication of magnetospheric emission would come from the detection of pulsations in the \textit{Fermi} emission of these sources whilst simultaneously confirming for the first time the nature of the compact object.

PSR B1259$-$63 is the only known millisecond pulsar in a binary system with a main-sequence star. It has been detected at TeV energies by H.E.S.S.~\cite{2005A&A...442....1A} during periastron passages of its 3.4 year orbital period.  \textit{Fermi} has yet to report any detection of pulsed magnetospheric emission from this source however, the upcoming periastron passage in December 2010 will offer the first opportunity for \textit{Fermi} to observe the expected GeV emission resulting from the pulsar wind interacting with the stellar wind of the companion.  Observations of this periastron passage will provide a key dataset to compare and contrast with those of the persistent $\gamma$-ray binaries in which the nature of the compact object is still unknown.

The detection of transient $\gamma$-ray activity from Cyg X-3 by both \textit{Fermi} and \emph{AGILE} during multiple epochs of soft-state transitions has confirmed this microquasar as a source of high energy emission.  Furthermore, it has been seen that the $\gamma$-ray emission is correlated with radio flare events.  The clear signature of orbital flux modulation detected by the LAT removes any doubt that Cyg X-3 is the source of the emission whilst also suggesting that the emitting region is likely bounded by the orbital separation of the system.

The recent reports by \emph{AGILE} of $\gamma$-ray activity from another microquasar, Cyg X-1, have not been confirmed by \textit{Fermi}.  The potential to discover flaring events from this source and other microquasars remains a real possibility through future observations with the LAT. Indeed with the upcoming periastron passage of PSR B1259$-$63, the detection of a symbiotic binary undergoing a nova outburst~\cite{ATel2487} and the ever present possibility of detecting a new $\gamma$-ray binary the future is very bright for \textit{Fermi}-LAT observations of binary systems.

\begin{acknowledgement}
The \textit{Fermi} LAT Collaboration acknowledges generous ongoing
support from a number of agencies and institutes that have supported
both the development and the operation of the LAT as well as
scientific data analysis.  These include the National Aeronautics and
Space Administration and the Department of Energy in the United
States, the Commissariat \`a l'Energie Atomique and the Centre
National de la Recherche Scientifique / Institut National de Physique
Nucl\'eaire et de Physique des Particules in France, the Agenzia
Spaziale Italiana and the Istituto Nazionale di Fisica Nucleare in
Italy, the Ministry of Education, Culture, Sports, Science and
Technology (MEXT), High Energy Accelerator Research Organization (KEK)
and Japan Aerospace Exploration Agency (JAXA) in Japan, and the
K.~A.~Wallenberg Foundation, the Swedish Research Council and the
Swedish National Space Board in Sweden.

Additional support for science analysis during the operations phase is
gratefully acknowledged from the Spanish CSIC and MICINN, the Istituto
Nazionale di Astrofisica in Italy and the Centre National d'\'Etudes
Spatiales in France.

ABH ackowledges funding by contract ERC-StG-200911 from the European Community.
\end{acknowledgement}

\end{document}